\documentclass[conference,a4paper]{IEEEtran}

\usepackage[utf8]{inputenc} 
\usepackage[T1]{fontenc}
\usepackage{url}
\usepackage{ifthen}
\usepackage{cite}
\usepackage[cmex10]{amsmath} 
                             \usepackage{amsfonts,amsmath,graphicx,subfig,algorithm}

\newcommand{\ZZ}{\mathbb{Z}}
\newcommand{\CC}{\mathbb{C}}

\newtheorem{proposition}{Proposition}

\begin{document}
\title{Compressed Neighbour Discovery using Sparse Kerdock Matrices} 


\author{%
  \IEEEauthorblockN{Andrew Thompson}
  \IEEEauthorblockA{Mathematical Institute\\
                    University of Oxford\\
                    Oxford, UK\\
                    Email: thompson@maths.ox.ac.uk}
  \and
  \IEEEauthorblockN{Robert Calderbank}
  \IEEEauthorblockA{Department of ECE\\
                    Duke University\\ 
                    Durham, NC\\
                    Email: robert.calderbank@duke.edu}
}


\maketitle

\begin{abstract}
We study the network-wide neighbour discovery problem in wireless networks in which each node in a network must discovery the network interface addresses (NIAs) of its neighbour. We work within the rapid on-off division duplex framework proposed by Guo and Zhang in~\cite{full_duplex} in which all nodes are assigned different on-off signatures which allow them listen to the transmissions of neighbouring nodes during their off slots; this leads to a compressed sensing problem at each node with a collapsed codebook determined by a given node's transmission signature.  We propose sparse Kerdock matrices as codebooks for the neighbour discovery problem. These matrices share the same row space as certain Delsarte-Goethals frames based upon Reed Muller codes, whilst at the same time being extremely sparse. We present numerical experiments using two different compressed sensing recovery algorithms, One Step Thresholding (OST) and Normalised Iterative Hard Thresholding (NIHT). For both algorithms, a higher proportion of neighbours are successfully identified using sparse Kerdock matrices compared to codebooks based on Reed Muller codes with random erasures as proposed in~\cite{dongning}. We argue that the improvement is due to the better interference cancellation properties of sparse Kerdock matrices when collapsed according to a given node's transmission signature. We show by explicit calculation that the coherence of the collapsed codebooks resulting from sparse Kerdock matrices remains near-optimal.
\end{abstract}


\section{Introduction}

In many wireless networks, such as mobile ad hoc networks (MANETs), each node can only communicate directly with a small number of other nodes, known as its neighbours. Before network-level activities are possible, each node must discover the network interface addresses (NIAs) of its neighbours. A node seeking to identify its neighbours (which we refer to as a query node) receives a linear superposition of the transmissions of its neighbours corrupted by noise (a multiple access channel), and then seeks to decode the NIAs of its neighbours. In state-of-the-art neighbour discovery protocols, the query node broadcasts a probe request and its neighbours reply by transmitting their NIAs repeatedly with random delays to ensure that they can be successfully retrieved with high probability despite collisions. 

It was argued in~\cite{dongning} that such protocols can be improved upon by assigning codewords to each node, which leads to a linear statistical inference problem. Since we may assume that the number of neighbours of a given query node is small, the problem becomes one of sparse recovery from linear measurements, also known as compressed sensing. 

Network-wide neighbour discovery brings with it a further challenge. Assuming each node is equipped with half-duplex hardware, nodes can either transmit or receive signals in a given time slot, but they cannot do both simultaneously. Naive solutions to this problem can introduce significant delays and waste channel resources. A clever way of achieving full-duplex communication using half-duplex radios, called random on-off division duplex (RODD), was proposed in~\cite{full_duplex}. In this approach, each user is assigned a unique on-off sequence, such that they are able to switch their radio to listening to other users' signals during their `off' slots. This approach was proposed in the context of neighbour discovery in~\cite{dongning}. 

A key question in this context is the design of a suitable codebook whose columns are the on-off sequences of the respective users. These codebooks must be amenable to effective and efficient decoding using compressed sensing algorithms, and furthermore they must be sparse. Luo and Guo~\cite{neighbor1,neighbor2} proposed using random Bernoulli codebooks along with a group testing reconstruction algorithm. These papers considered neighbour discovery at a single query node rather than network-wide neighbour discovery. The authors extended the approach to network-wide neighbour discovery in~\cite{dongning} and also proposed the use of codebooks based on Reed-Muller codes (Delsarte-Goethals frames; see Section~\ref{kerdock}) with random erasures. They showed that the Reed-Muller based approach led to improvements in reconstruction performance when used in conjunction with a modified chirp reconstruction algorithm~\cite{chirp}.

Rather than achieving sparsity by making erasures to existing codebook designs, in this paper we consider codebooks which are sparse by design. In particular, we consider a family of sparse matrices, which we refer to as \textit{sparse Kerdock matrices} which combine two useful properties: they are extremely sparse, and they share the same row space as certain Delsarte-Goethals (DG) frames. Delsarte-Goethals frames~\cite{DG} are a popular choice as a codebook for neighbour discovery, and as measurement matrices for compressed sensing more generally, due to their near-optimal coherence properties. Since sparse Kerdock matrices share the same row-space, their Gram matrix is precisely the same as the Delsarte-Goethals frame to which they are related. The sparsity property, on the other hand, ensures that each user transmits during, and therefore misses, only $n$ out of $n^2$ time slots, which means that the Gram matrix and the excellent coherent properties are preserved.

We demonstrate that the performance of compressed neighbour discovery is significantly improved when using a sparse Kerdock matrix codebook as compared to a standard Delsarte-Goethals frame with erasures. We perform numerical tests upon a network propagation loss model introduced by Zhang and Guo~\cite{dongning} and in the context of two reconstruction algorithms. The first algorithm is the well-known One Step Thresholding algorithm~\cite{OST}, which is also known as maximum likelihood detection and TIN (`Treat Interference as Noise')~\cite{polyanskiy}. We also consider a more powerful but more computationally demanding iterative algorithm called Normalised Iterative Hard Thresholding (NIHT)~\cite{NIHT} which is a popular choice in the compressed sensing community. We also provide theoretical justification for the improvement in performance by calculating the coherence of the collapsed codebook restricted to the rows (time slots) for which a given user is receiving data.

The structure of the rest of the paper is as follows. In Section~\ref{model} we describe a model for the neighbour discovery problem and network propagation loss presented first in~\cite{dongning}. In Section~\ref{kerdock} we present the sparse Kerdock matrix construction first introduced in~\cite{globalsip}. In Section~\ref{experiments} we present the results of our numerical experiments on the neighbour discovery problem, and in Section~\ref{coherence} we give a theoretical result on the coherence of the reduced codebooks resulting from sparse Kerdock matrices and present coherence statistics for both sparse Kerdock matrices and codebooks based upon random erasures, before concluding in Section~\ref{conclusion}.

\section{Channel and network models}\label{model}

We briefly describe the channel and network models considered in~\cite{dongning}. 

\subsection{Channel model}

We assume a network of $N$ nodes, and a neighbour discovery interval of $M$ symbols so that each node is assigned a length-$M$ codeword. We also assume symbol-synchronous transmission between all nodes. We write $a_i$ for the codeword assigned to node $i$, and $A=\begin{bmatrix}a_1&a_2&\ldots&a_N\end{bmatrix}$ for the full codebook matrix. Writing $\mathcal{N}_q$ for the neighbours of node $q$, we model the received signal $y^q$ at node $q$ as the linear superposition of the signals transmitted by its neighbours, that is
$$y^q=\sqrt{\gamma}\sum_{i\in\mathcal{N}_q}x_i a_i + w^q,$$
where $x_i$ is the complex-valued wireless coefficient of the link between node $q$ and node $i$, the entries of $w^q$ are i.i.d. $\mathcal{CN}(0,1)$ random variables and $\gamma$ is the average channel gain in the SNR. Any signal received from non-neighbouring nodes is accounted for by the noise. Therefore, writing $x^q=\begin{bmatrix}x_1&x_2&\ldots&x_N\end{bmatrix}^T$ where $x_i=0$ if $i$ is not a neighbour of node $q$, we can equivalently write
$$y^q=\sqrt{\gamma}Ax^q + w^q.$$
To enable network-wide neighbour discovery, we assume that each node's codeword is a sparse on-off signature, and that the node only receives signals during its off slots. Let $\bar{y^q}$ and $\bar{w^q}$ denote the measurements and noise respectively received by node $q$ during its off slots, and denote by $A^q$ the collapsed codebook matrix which is the matrix $A$ with rows restricted to the off slots of node $q$ removed. Then the neighbour discovery problem at node $q$ consists in identifying the nonzero coefficents of the sparse vector $x^q$ from the linear measurements
$$\bar{y^q}=\sqrt{\gamma}A^q x^q+\bar{w^q}.$$

\subsection{Network model}

Starting from the assumption that all nodes are distributed in a plane according to a homogeneous Poisson process, and that the channel power gain between pairs of nodes decays according to a power law, the following model was derived in~\cite{dongning} for the coefficients $x_i$. Write $x_i=b_i u_i$, where $b_i\sim Bin(N,k/N)$ is a Binomial random variable with expectation $k$ which determines the neighbours, and where $u_i$ is a random variable with pdf
$$f_{u_i}(u)=\left\{\begin{array}{ll}
\displaystyle\frac{4}{\alpha}\cdot\frac{\eta^{2/\alpha}}{u^{4/\alpha+1}}&u\geq\sqrt{\eta}\vspace{0.2cm}\\
0&\textrm{otherwise.}\end{array}\right.$$
Here $\alpha$ and $\eta$ are parameters: $\alpha$ is the exponent in the power law decay and $\eta$ is a threshold which determines whether or not a node is deemed to be a neighbour.

\section{Sparse Kerdock matrices}\label{kerdock}

Sparse Kerdock matrices were defined in~\cite{globalsip} as follows. Given a positive integer $m$, index the rows of a $2^{2m}\times 2^{3m}$ matrix $S^m$ by $(u_1,u_2)$ and its columns by $(a_1,a_2,b)$ where $u_1,u_2,a_1,a_2,b\in\ZZ_2^m$ are each binary $m$-tuples. Define its entries to be
\begin{equation}\label{formula}S^m_{(u_1,u_2),(a_1,a_2,b)}=\left\{\begin{array}{ll}\displaystyle\frac{1}{2^{m/2}}(-1)^{u_1 a_1^T}&u_1 P_b+u_2=a_2\vspace{0.2cm}\\
0&\textrm{otherwise},\end{array}\right.
\end{equation}
where $\{P_b\}$ is a \emph{Kerdock} set of binary symmetric matrices which has the property that the sum of all distinct elements is full rank. Sparse Kerdock matrices are indeed extremely sparse, as the following result from~\cite{globalsip} establishes.

\begin{proposition}{\cite[Proposition 2]{globalsip}}\label{sparsity}
Each row of $S^m$ has $2^{2m}$ nonzeros and each column of $S^m$ has $2^m$ nonzeros.
\end{proposition}

Writing $n=2^m$, $S^m$ is of size $n^2\times n^3$, and every column has $n$ nonzero coefficents. If the columns are used as codewords for rapid on-off division duplex, each node is able to receive signals during $(n^2-n)$ of the $n^2$ symbol slots.

Furthermore, it was established in~\cite[Proposition 1]{globalsip} that sparse Kerdock matrices share the same row space as certain incomplete Delsarte-Goethals (DG)~\cite{DG} frames of type $DG(2m,0)$. Indeed, there exists a unitary transformation which can be applied to the rows of a sparse Kerdock matrix to obtain an incomplete Delsarte-Goethals frame and vice versa. Delsarte-Goethals frames are closely related to second-order Reed-Muller codes over $\ZZ_4$: more precisely, the columns of Delsarte-Goethals frames are obtained by exponentiating Reed-Muller codewords. 

DG frames are popular choices for measurement matrices in compressed sensing due to their excellent coherence properties. Given a matrix $A=\begin{pmatrix} a_1&a_2&\ldots&a_N\end{pmatrix}$ with $N$ nonzero columns, define its \emph{coherence} $\mu(A)$ to be
$$\mu(A):=\max_{i\neq j}\frac{|a_i^*a_j|}{\|a_i\|_2\|a_j\|_2}.$$
It is well-known that DG frames of type $DG(2m,0)$ have coherence $1/2^m$~\cite{DG}, which is close to the optimal Welch bound~\cite{welch}. Since sparse Kerdock matrices are obtained from $DG(2m,0)$ frames by unitary transformation, they share the same Gram matrix and have the same coherence. Therefore, the $n^2\times n^3$ sparse Kerdock matrix $S^m$ has coherence $1/n$.

\begin{figure}[h!]
\begin{center}
\includegraphics[width=0.4\textwidth]{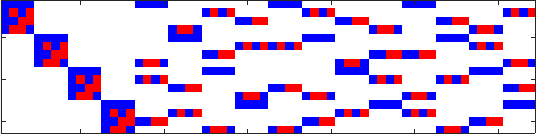}
\caption{Example of a sparse Kerdock matrix $S^m$ for $m=2$. Blue $+$; red $-$; white $0$.}\label{sparse_kerdock_examples}
\end{center}
\end{figure}

Figure~\ref{sparse_kerdock_examples} gives an example of a sparse Kerdock matrix for the case $m=2$ ($16\times 64$). 

We can note that columns are either orthogonal, or else there is an overlap of at most one between the components for which the coefficients of any pair of columns are nonzero. In other words, sparse Kerdock matrices are a highly structured approach to on-off division duplex. It is this structure which means that coherence properties are preserved when the codebook matrix is collapsed according to the on-off signatures of a given query node. A key message of this paper is that designing structured sparse codebooks is preferable to random erasures of existing codebooks for the network-wide neighbour discovery problem.

\section{Numerical experiments}\label{experiments}

In this section, we compare the use of sparse Kerdock matrices with DG frames (Reed-Muller codes) with random erasures. More precisely, we construct an incomplete $DG(2m,0)$ frame $K^m$ of the same dimensions as $S^m$, namely $2^{2m}\times 2^{3m}$ or $n^2\times n^3$. We then perform pointwise multiplication of $K^m$ by a Bernoulli random mask $B^m$ whose entries are i.i.d. $1$ with probability $1/2^r$ and $0$ with probability $(1-1/2^r)$. The parameter $r$, which determines the proportion of on slots versus off slots in the signatures, can be varied. In our numerical experiments we optimize over integer values of $r$ in $\{1,2,3,4\}$. This construction -- DG frames with random erasures -- is the same in priciple as the one proposed in ~\cite{dongning}, which considers masks which are not completely i.i.d. random but consist of smaller random masks replicated several times. It is worth noting that structured random erasures are appealing from the point of view of efficient coding of the random patterns. 

We perform two experiments: one with the One Step Thresholding (OST) algorithm and the other with the Normalised Iterative Hard Thresholding (NIHT) algorithm.

OST~\cite{OST} is a popular low-complexity approach to signal recovery in compressed sensing in which the measurements are projected back onto the row space of the measurements and the largest in magnitude coefficients are taken as a predictor of the nonzero coefficients. The algorithm is summarized in Algorithm~\ref{OST}.
\begin{algorithm}
\textbf{Inputs}: $\bar{y^q}\in\CC^{(n^2-n)}$, $A^q\in\CC^{(n^2-n)\times n^3}$, $s\in\ZZ_+$.
\begin{enumerate}
\item $g^q=(A^q)^*\bar{y^q}$.
\item $\Gamma^q:=\{i\;\textrm{corresponding to the}\;s\;\textrm{largest}\;|g^q_i|\}.$
\end{enumerate}
\textbf{Outputs}: $\Gamma^q$.
\caption{One Step Thresholding}
\label{OST}
\end{algorithm}
Note that OST requires an input parameter $s$ which is the number of nonzero coefficients sought, and it outputs an index set $\Gamma^q$ corresponding to the predicted locations of the nonzero coefficients (neighbours).

\begin{figure}[h!]
\begin{center}
\includegraphics[width=0.4\textwidth]{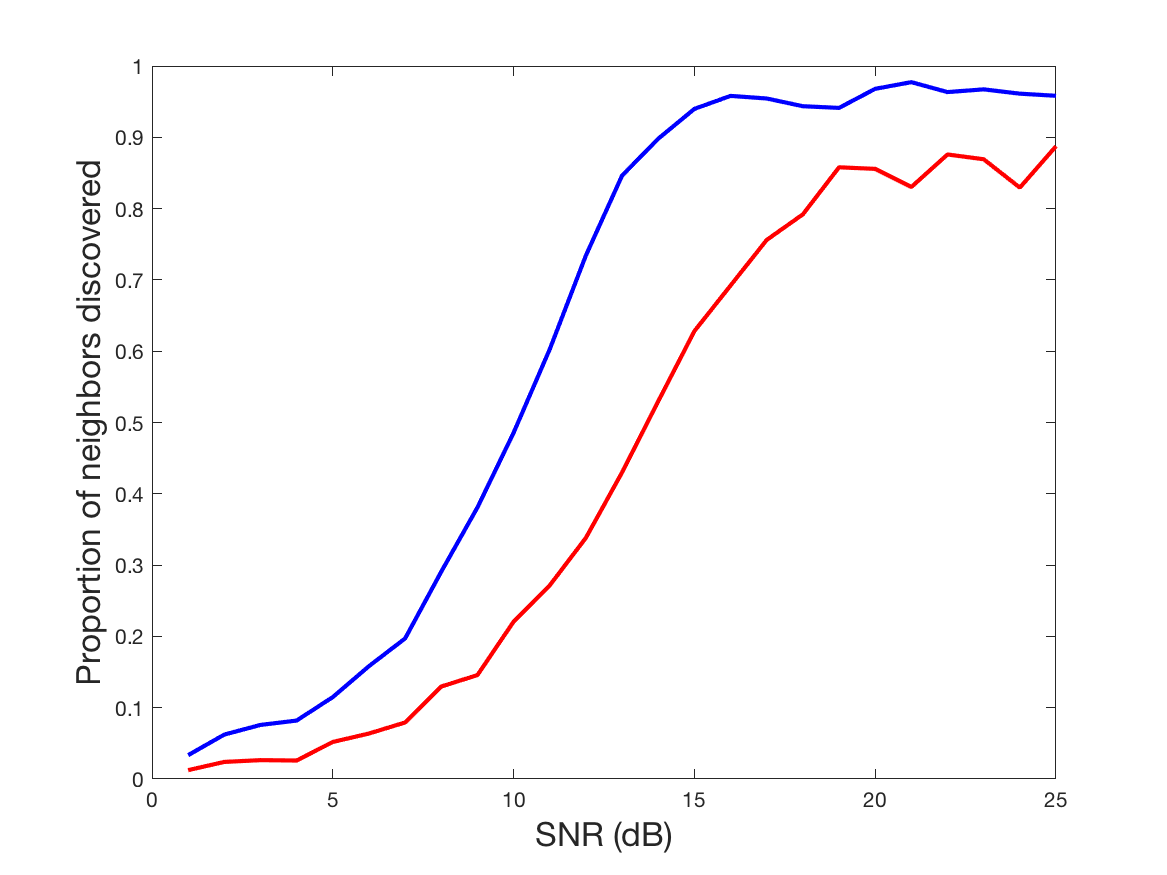}
\caption{Proportion of discovered neighbours against SNR using OST; blue: sparse Kerdock matrices; red: DG frames with random erasures.}\label{OST_plot}
\end{center}
\end{figure}

We generate $200$ random instances of the system model described in Section~\ref{model}. We assume there are $N=2^{15}$ nodes in our network and take $m=5$ ($n=2^5$) to give codebook matrices of size $2^{10}\times 2^{15}$ ($M=2^{10}$). We take $\eta=0.05$ and $\alpha=3$ in the propagation model and set the average number of neighbours per node to be $k=5$. We record the proportion of neighbours successfully identified by the OST algorithm over all $200$ nodes. We assume that some prior knowledge is available in advance concerning an upper bound on the number of neighbours of a given node. In line with this assumption, we set the parameter $s$ in OST to be $3k=15$, though we note in passing that the results are not particularly sensitive to the choice of $s$ providing it exceeds $k$. Figure~\ref{OST_plot} plots the proportion of successfully discovered neighbours against SNR (dB) for both codebook designs. We observe a significant increase in the proportion of discovered neighbours using sparse Kerdock matrices compared to DG frames with random erasures. OST is intuitively treating interference as noise: we argue in Section~\ref{coherence} that the reason for the improved performance is that the interference cancellation properties of sparse Kerdock matrices are better preserved when collapsed according to the on-off signature of a given query node, which in turn is due to the structured nature of the on-off signatures themselves.

We perform a further experiment using a different reconstruction algorithm in order to demonstrate that the observed improvement is not algorithm-specific. Normalised Iterative Hard Thresholding (NIHT) is a more computationally intensive but better performing compressed sensing reconstruction algorithm which can be viewed as an iterative extension of OST in which the signal approximation is repeatedly projected onto the row space of the measurements and thresholded. The algorithm requires the same input parameters as OST and outputs a signal approximation which may be thresholded to give a prediction for the locations of the nonzero coefficients. We refer the reader to~\cite{NIHT} for further details on the algorithm.

\begin{figure}[h!]
\begin{center}
\includegraphics[width=0.4\textwidth]{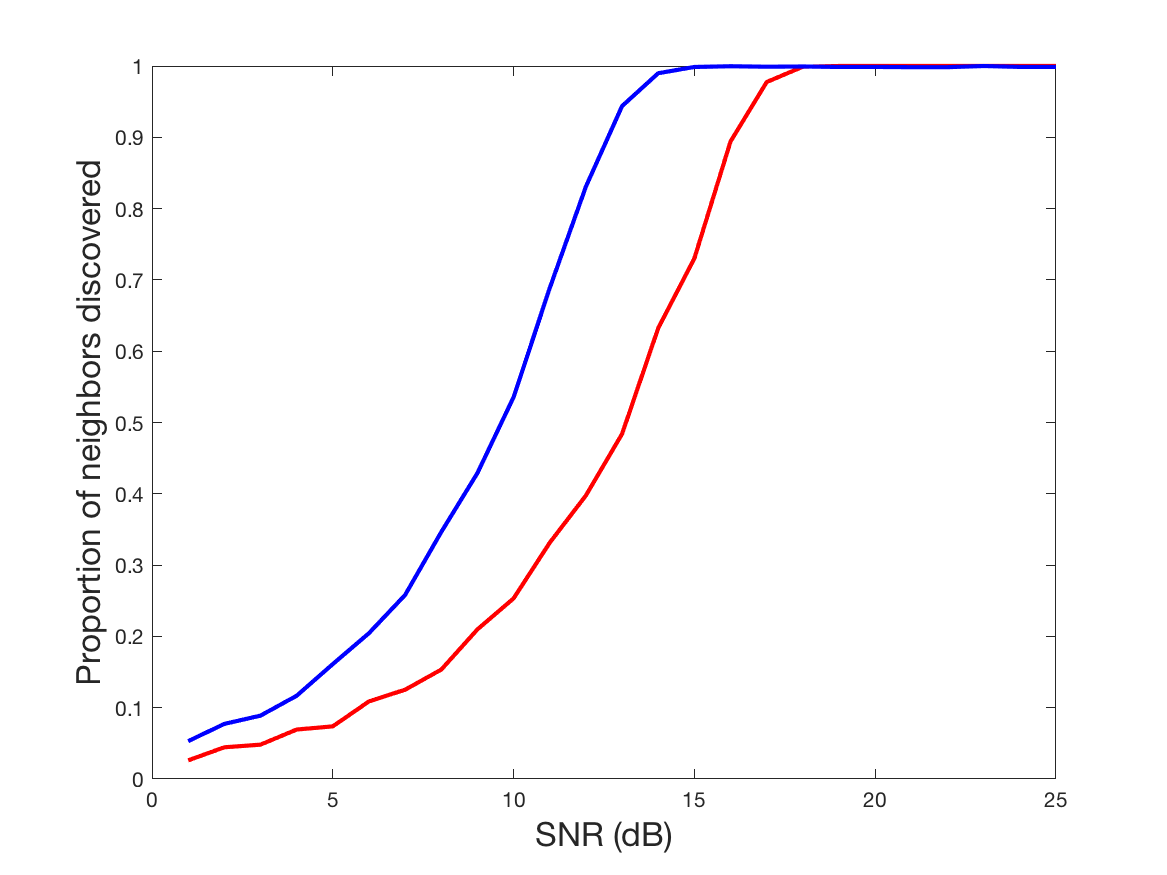}
\caption{Proportion of discovered neighbours against SNR using NIHT; blue: sparse Kerdock matrices; red: DG frames with random erasures.}\label{NIHT_plot}
\end{center}
\end{figure}

We generate $50$ random instances of the system model described in Section~\ref{model}, again taking $M=2^{10}$, $N=2^{15}$, $\eta=0.05$ and $\alpha=3$, but this time setting the average number of neighbours per node to be $k=60$. We record the proportion of neighbours successfully identified by the NIHT algorithm over all $50$ nodes, this time setting the parameter $s$ in OST to be $3k=180$. Figure~\ref{NIHT_plot} plots the proportion of successfully discovered neighbours against SNR (dB) for both codebook designs. We again observe a significant increase in the proportion of discovered neighbours using sparse Kerdock matrices compared to DG frames with random erasures. We note in addition that the SNR threshold above which essentially all neighbours are successfully identified decreases from around $18$ for DG frames with random erasures to around $15$ for sparse Kerdock matrices.

\section{Coherence of collapsed codebooks}\label{coherence}

Given a matrix $A=\begin{bmatrix} a_1&a_2&\ldots&a_p\end{bmatrix}\in\CC^{M\times N}$ with nonzero columns, define its \emph{coherence} $\mu(A)$ to be
$$\mu(A):=\max_{i\neq j}\frac{|a_i^*a_j|}{\|a_i\|_2\|a_j\|_2}.$$

Recalling (\ref{formula}), fix the query node $q$ to be the one indexed by $(a_1^*,a_2^*,b^*)$. This node is blind to the measurements corresponding to the rows indexed by $\{(u_1,u_2)\,:\,u_1 P_{b^*}+u_2=a_2^*\}$. Note also that a node indexed by $(a_1,a_2^*,b^*)$ for any $a_1$ transmits in the same rows, from which it follows that the query node is completely blind to these nodes, of which there are $n$ of them (which includes itself). Removing the rows indexed by $\{(u_1,u_2)\,:\,u_1 P_{b^*}+u_2=a_2^*\}$ and the columns indexed by $\{(a_1,a_2,b)\,:\,a_2=a_2^*,b=b^*\}$, we obtain a collapsed codebook matrix $S^m_q$ of size $(n^2-n)\times(n^3-n)$. The next result shows that the coherence of $S^m_q$ is near-optimal.

\begin{proposition}
The collapsed codebook matrix $S^m_q$ with $p=n^2-n$ rows has coherence
$$\mu(S^m_q)=\frac{2}{\sqrt{4p+1}-1}\approx\frac{1}{\sqrt{p}}.$$
\end{proposition}

\emph{Proof}: We first show that any two distinct columns of $S^m$ are either orthogonal or have a single nonzero component in common. Consider two columns indexed by $(a_1^1,a_2^1,b^1)$ and $(a_1^2,a_2^2,b^2)$ respectively. Overlapping nonzero entries must simultaneously satisfy
\begin{equation}\label{eqn1}
u_1 P_{b^1}+u_2=a_2^1
\end{equation}
and
\begin{equation}\label{eqn2}
u_1 P_{b^2}+u_2=a_2^2.
\end{equation}
Adding the two equations, we obtain
\begin{equation}\label{eqn_add}
u_1(P_{b^1}+P_{b^2})=a_2^1+a_2^2.
\end{equation}
First suppose $b^1=b^2$. Then (\ref{eqn_add}) implies that $a_2^1=a_2^2$, which combines with (\ref{formula}) to give
$$\begin{array}{l}(S^m_{(u_1,u_2),(a_1,a_2,b)})^*S^m_{(u_1,u_2),(a_1^2,a_2^2,b^2)}\\\;\;\;\;=\displaystyle\sum_{\{(u_1,u_2):u_1P_{b_1}+u_2=a_2^1\}}(-1)^{u_1 (a_1^2-a_1^1)^T}\\\;\;\;\;\;\;\;\;\;\;\;\;\;\;\;\;\;\;\;\;=\displaystyle\sum_{u_1}(-1)^{u_1 (a_1^2-a_1^1)^T}.\end{array}$$
Since the columns are distinct, $a_1^1\neq a_1^2$, and it follows by standard properties of Hadamard sums that the columns are orthogonal. Now suppose $b_1\neq b_2$. Then the properties of Kerdock sets imply that $P_{b^1}+P_{b^2}$ is full rank. There exists therefore a unique solution $u_1$ to (\ref{eqn_add}), and it is easy to then deduce that there exists a unique solution $(u_1,u_2)$ simultaneously satisfying (\ref{eqn1}) and (\ref{eqn2}), and so the two columns have a single nonzero component in common. We next show that precisely one entry is erased from each remaining column of $S^m_q$. Without loss of generality, consider the column indexed by $(a_1',a_2',b')$. An entry indexed by $(u_1,u_2)$ is erased if it simultaneously satisfies
\begin{equation}\label{eqn1v2}
u_1 P_{b'}+u_2=a_2'
\end{equation}
and
\begin{equation}\label{eqn2v2}
u_1 P_{b^*}+u_2=a_2^*.
\end{equation}
Adding the two equations, we obtain
\begin{equation}\label{eqn_addv2}
u_1(P_{b'}+P_{b^*})=a_2'+a_2^*.
\end{equation}
First suppose $b'=b^*$, in which case (\ref{eqn_addv2}) implies that $a_2'=a_2^*$, which contradicts the assumption that all such columns have been removed from $S^m$. It follows that $b'\neq b^*$, and an argument analogous to the one just above then establishes that there exists a unique solution $(u_1,u_2)$ simultaneously satisfying (\ref{eqn1v2}) and (\ref{eqn2v2}), and so precisely one entry is erased from each remaining column of $S^m_q$. Next we show that if two columns of $S^m$ are orthogonal, the nonzero entry removed from each column is in the same position. Supposing the orthogonal columns to be indexed by $(a_1^1,a_2,b)$ and $(a_1^2,a_2,b)$ respectively, nonzero components removed from either column must simultaneously satisfy
\begin{equation}\label{eqn1v3}
u_1 P_{b}+u_2=a_2
\end{equation}
and
\begin{equation}\label{eqn2v3}
u_1 P_{b^*}+u_2=a_2^*,
\end{equation}
and an analogous argument to the one above then establishes that there exists a unique solution $(u_1,u_2)$ simultaneously satisfying (\ref{eqn1v3}) and (\ref{eqn2v3}), and so the nonzero entry removed from each column is in the same position. We can now calculate all possible absolute values of inner products of pairs of columns of the collapsed matrix $S^m_q$. If the columns of $S^m$ are orthogonal, the corresponding columns $s_1$ and $s_2$ of $S^m_q$ satisfy $\|s_1\|_2^2=\|s_2\|_2^2=(n-1)/n$ and $|s_1^*s_2|=1/n$, which gives
$$\frac{|s_1^*s_2|}{\|s_1\|_2\|s_2\|_2}=\frac{1}{n-1}.$$
Meanwhile, if the columns of $S^m$ are not orthogonal, either the entries in their shared nonzero component are removed, making $s_1$ and $s_2$, the corresponding columns of $S^m_q$, orthogonal, or else nonzero entries are removed from nonoverlapping components, and so 
$$\frac{|s_1^*s_2|}{\|s_1\|_2\|s_2\|_2}=\frac{1}{n}$$
as before. It follows that the coherence of $S^m_q$ is equal to $\frac{1}{n-1}$. The result now follows on substituting $p=n^2-n$ or equivalently $n=(\sqrt{4p+1}+1)/2$.\hfill$\Box$

Next we numerically compare the coherence of the collapsed matrices arising from sparse Kerdock matrices to the coherence of the collapsed matrices resulting from DG frames with erasures. We take $m=4$ and matrices of size $2^8\times 2^{12}$. For DG frames, we vary the parameter $r$ over $\{1,2,3,4\}$. Recall that the proportion of erasures is $(1-1/2^r)$. Table~\ref{coherences} gives the coherence, averaged over $10$ independent trials in the case of DG frames with erasures. We observe that near-optimal coherence is not preserved when DG frames with erasures are collapsed for a given query node. The explanation for the increased coherence is either that there is significant overlap between the locations of the nonzeros of different columns leading to greater information loss and disruption to the Gram matrix.

\begin{table}
\centering
\begin{tabular}{|c|c|cccc|}
\hline
&Sparse&&DG&Frames&\\
&Kerdock&$r=1$&$r=2$&$r=3$&$r=4$\\
\hline
Coherence&$0.0667$&$0.4530$&$0.3800$&$0.3886$&$0.4784$\\
\hline
\end{tabular}
\caption{Average coherence of `collapsed' matrices arising from sparse Kerdock matrices and DG frames with erasures.}\label{coherences}
\end{table}

\begin{figure}[h!]
\begin{center}
\includegraphics[width=0.4\textwidth]{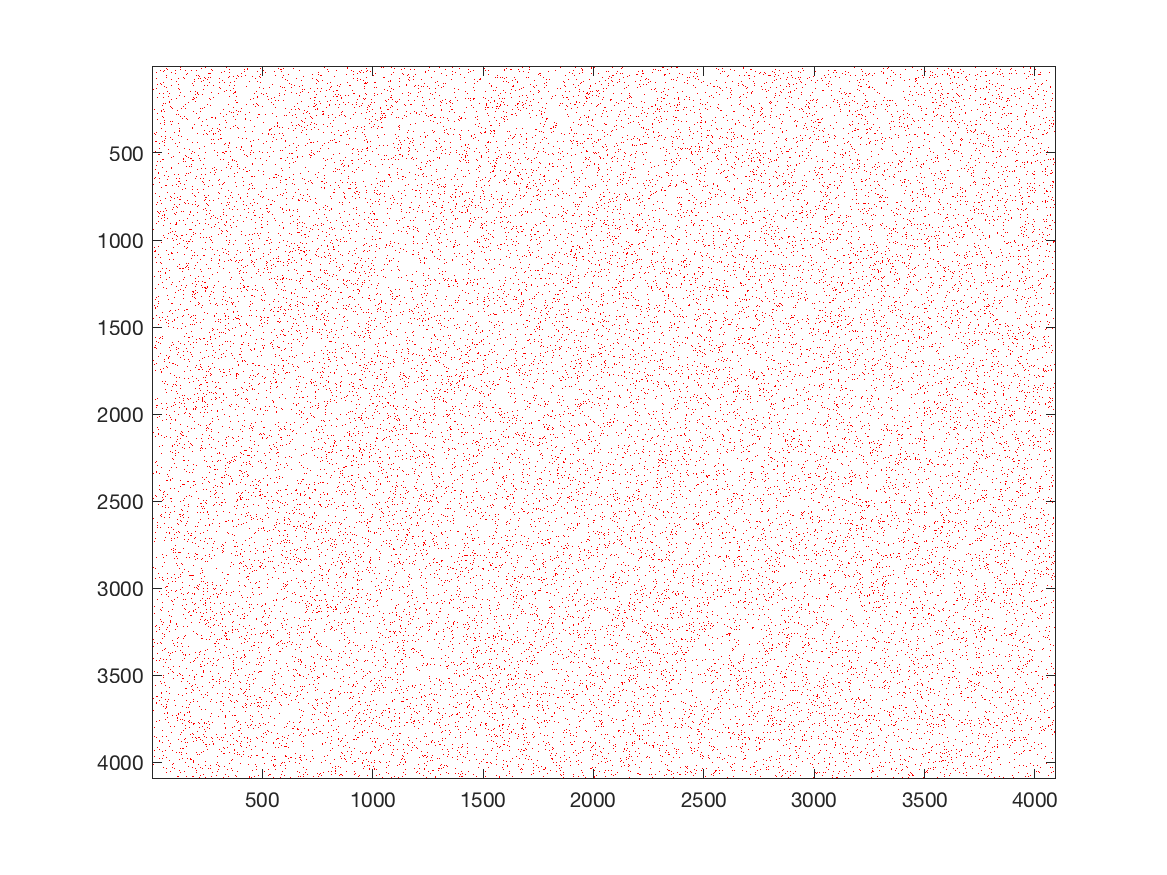}
\caption{Large entries (greater than $0.15$ in red) of the normalized Gram matrix of the collapsed matrix arising from DG frames with erasures.}\label{DG_gram}
\end{center}
\end{figure}

We further illustrate the comparison by plotting an example of the Gram matrix for for the DG frame with erasure construction. Since it has the smallest coherence, we choose $r=2$. We first normalize the columns of the collapsed measurement matrix $A^q$ to give $\bar{A^q}$ and then plot in red in Figure~\ref{DG_gram} the entries of the Gram matrix $\bar{A^q}^*\bar{A^q}$ which have absolute value greater than $0.15$. Figure~\ref{DG_gram} illustrates that in the case of DG frames there are a significant number of entries in the Gram matrix with absolute value above $0.15$. Distinguishing between column pairs corresponding to these entries will be somewhat difficult. We ignore the diagonal of the Gram matrix since these entries give column norms and are not relevant to the coherence calculation.

One caveat to note is that, while sparse Kerdock matrices have smaller coherence, they do have the drawback that any given node will be completely blind to a small fraction of the other nodes and will therefore be unable to detect them. Given a query node indexed by $(a_1^*,a_2^*,b^*)$, this happens for nodes indexed by $(a_1,a_2,b)$ such that $a_2=a_2^*$ and $b=b^*$. More precisely, $n-1$ of the $n^3-1$ other nodes are undetectable. For $n=16$ as considered in this section, this amounts to a fraction of $1/273$. One way to mitigate this effect would be a judicious assignment of the codewords throughout the network. 

\section{Conclusions and future directions}\label{conclusion}

We have proposed sparse Kerdock matrices for network-wide neighbour discovery and demonstrated that they lead to improved neighbour detection rates compared to previous constructions based on DG frames. We have also provided theoretical justification for the improved performance by analysing the coherence properties of the collapsed measurement matrices resulting from the on-off signature of a given query node.

We presented results for two compressed sensing algorithms: OST and NIHT. Both of these algorithms require computations at least $\mathcal{O}(N)$ in the number of nodes $N$, which becomes too computationally demanding for the large $N$ expected in practice (e.g. $N=2^{48}$~\cite{dongning}). A more computationally efficient algorithm is the chirp reconstruction algorithm~\cite{chirp}, which has sublinear complexity in $N$. Since sparse Kerdock matrices are related to DG frames by unitary transformation, it is clear that the chirp reconstruction algorithm can also be used in conjunction with sparse Kerdock matrices. In future work, we plan to extend the chirp reconstruction algorithm for use with sparse Kerdock matrices. Indeed, a version of the chirp reconstruction algorithm adapted to the case of DG frames with erasures was proposed in~\cite{dongning}. It is worth noting that the performance improvements over DG frames with erasures would be expected to also be observed in conjunction with chirp reconstruction, since in the sparse Kerdock case the original algorithm can be used without even the need to adapt it to deal with erasures.

Another question that naturally arises is whether there exist other sparse matrix constructions with similar properties to sparse Kerdock matrices. In fact there do exist other families of matrices with similar coherence and sparsity properties which would also be expected to perform well in the context of neighbour discovery. Examples of such constructions are certain types of Steiner equiangular tight frames (ETFs)~\cite{steiner}. However, sparse Kerdock matrices have two obvious advantages over Steiner ETFs: their matrix-vector products can be computed using fast transforms based upon the Walsh-Hadamard transform (unlike Steiner ETFs) and also sparse Kerdock matrices are amenable to decoding using the chirp reconstruction algorithm (unlike Steiner ETFs).

We also note that the ratio between the length of the codewords ($n^2$) and the number of nodes ($n^3$) also somewhat limits the number of nodes that can be included in the network. An interesting area for future exploration would be to design matrices with smaller aspect ratio but which still have a certain degree of structured sparsity which leads to improved performance in the context of neighbour discovery.




\end{document}